\documentclass[%
twocolumn,
reprint,
preprintnumbers,
amsmath,amssymb,
API
]{revtex4-2}
\UseRawInputEncoding
\usepackage{graphicx}% Include figure files
\usepackage{dcolumn}% Align table columns on decimal point
\usepackage{bm}% bold math
\usepackage{amssymb}
\usepackage{amsmath}
\usepackage{color}
\usepackage{soul}
\usepackage{subfig}
\usepackage{hyperref}% add hypertext capabilities

\begin{document}

\title{Vacuum Rabi Splitting and Quantum Fisher Information of a Non-Hermitian Qubit in a Single-Mode Cavity		}
\author{Yi-Cheng Wang}
\author{Jiong Li}
\affiliation{Zhejiang Key Laboratory of Micro-Nano Quantum Chips and Quantum Control, School of Physics, Zhejiang University, Hangzhou 310027, China }
\author{Li-Wei Duan}
\email{duanlw@zjnu.edu.cn}
\affiliation{Department of Physics, Zhejiang Normal University, Jinhua 321004, China }
\author{Qing-Hu Chen}
\email{qhchen@zju.edu.cn}
\affiliation{Zhejiang Key Laboratory of Micro-Nano Quantum Chips and Quantum Control, School of Physics, Zhejiang University, Hangzhou 310027, China }
\affiliation{Collaborative Innovation Center of Advanced Microstructures, Nanjing
University, Nanjing 210093, China }
\date{\today}

\begin{abstract}
A natural extension of the non-Hermitian qubit is to place it in a single-mode cavity. This setup corresponds to the quantum Rabi model (QRM) with a purely imaginary bias on the qubit, exhibiting  parity-time ($\mathcal{P}\mathcal{T}$)  symmetry. In this work, we first solve the $\mathcal{P} \mathcal{T}$-symmetric QRM  using the Bogoliubov operator approach.  We  derive the transcendental function responsible for the exact solution, which can also be used to precisely identify  exceptional points. The adiabatic approximation previously used can be easily formulated within this approach by considering transitions between the same manifolds  in the space of  Bogoliubov operators.	By further considering transitions between the nearest-neighboring manifolds, we can analytically obtain more accurate eigensolutions.	 Moreover, these simple corrections can capture the main features of the dynamics, where the adiabatic approximation fails. Furthermore, the rich characteristics  of the vacuum Rabi splitting in the emission spectrum  are predicted. The width of the peaks  increases with the  coupling strength and the imaginary biases, reflecting the nature of  open quantum systems. Additionally, we identify a {quantum-criticality-enhanced} effect by calculating the quantum Fisher information. Near the exceptional points, the quantum Fisher information in the $\mathcal{P} \mathcal{T}$-symmetric QRM is significantly higher than that of the non-Hermitian qubit component.	 This may open a new avenue for enhancing quantum sensitivity in non-Hermitian systems by incorporating coupling with an additional degree of freedom, enabling more precise parameter estimation.		
\end{abstract}

%\keywords{non-Hermitian quantum Rabi models, the Bogoliubov operators approach, exceptional points }
\maketitle

\section{Introduction}

In recent years, concepts from non-Hermitian physics and parity-time ($\mathcal{P}$$\mathcal{T}$) symmetry have attracted considerable
interest. Bender et al. demonstrated that a non-Hermitian Hamiltonian
invariant under combined space reflection and time
reversal  can exhibit a real spectrum \cite{Bender1998}. This
counterintuitive result challenges the traditional view that real
eigenvalues are associated exclusively with Hermitian observables, marking a
significant milestone in non-Hermitian physics \cite%
{heiss2004,Mostafazadeh2004,Bender2007,ashida2020,bender2023}.

Unlike the strict mathematical requirement in traditional quantum physics
that a Hamiltonian operator should be Hermitian, the $\mathcal{PT}$
invariance condition is a weaker, more physically intuitive constraint,
providing a rich field for both theoretical and experimental exploration. As
system parameters vary, the $\mathcal{PT}$-symmetric system can
transition between the $\mathcal{PT}$-unbroken and $\mathcal{P}%
\mathcal{T}$-broken phases. In the $\mathcal{PT}$-unbroken phase,
the system has real eigenvalues, while in the $\mathcal{PT}$%
-broken phase, it has a pair of conjugate eigenvalues. Phase transitions are
characterized by numerous accumulation points of exceptional points (EPs).
At these EPs, both eigenvectors and eigenvalues coalesce. Additionally,
dimensionality reduction occurs in the eigenspace of the EPs \cite%
{el-ganainy2018,meden2023}.

Non-Hermitian $\mathcal{PT}$ symmetry has recently received considerable attention in the study of $\mathcal{PT}$-symmetric systems involving the interaction of qubits with classical light fields~\cite{jogleka2014,leePT2015,xieTime2018,jogleka2014,Liu2024}. This research belongs to the context of the non-Hermitian semiclassical Rabi model.	$\mathcal{PT}$\-symmetry in pure quantum systems, particularly in the context of light-matter interaction with an  imaginary qubit-cavity coupling, has also been studied. Examples include the non-Hermitian Jaynes-Cummings model \cite{NJC2020}, the non-Hermitian double Jaynes-Cummings model \cite{LiJC2023}, the dissipative quantum Rabi model (QRM) \cite{LeeCh2021}, and the non-Hermitian QRM \cite{LiNR2025}.	

The $\mathcal{PT}$-symmetric quantum Rabi model (PTQRM) with a purely imaginary bias \cite{LiPT2023} has rarely been explored. This model is, however,  effectively  a natural extension of the well-studied non-Hermitian two-level system (NHTLS) coupled to a quantized light field.	The PTQRM exhibits intriguing non-Hermitian properties, including an infinite number of EPs that vanish and reappear depending on the strength of the light-matter coupling.	Furthermore, the PTQRM can be realized in  realistic physical systems \cite{LiPT2023}.	Potential physical implementations of the non-Hermitian Hamiltonian for the PTQRM  include circuit quantum electrodynamics (cQED) \cite{naghiloo2019}, trapped-ion experiments, and the replacement of traditional qubits in QRM setups \cite{naghiloo2019, yoshihara2017, trappedion2021, ioncavity2023, cai2021}.	

In this work, we extend the Bogoliubov operator approach (BOA) \cite%
{qhchen2012,DuanMix2019,xiePRR2021} to  solve the PTQRM \cite%
{LiPT2023}, and obtain the exact eigensolutions and the EPs. Within this framework, we  analytically propose a correction  to the previous adiabatic approximation (AA), especially the eigenstates, leading to a deeper understanding of dynamics.  We primarily investigate two prominent quantum effects of this model using our developed approaches: vacuum Rabi splitting (VRS), a key phenomenon in atomic physics ~\cite{sanchez1983}, and  Quantum Fisher information (QFI), which plays a crucial role in quantum metrology.  Recent proposals suggest that quantum sensors based on non-Hermitian systems may exhibit enhanced sensitivity	
~\cite{QuantumMetrology2011,CriticalPara2023,NingboQFI2023,ying2024,liuzh2016,yu2020,liao2021,ding2023,wang2024}%
. We calculate the QFI for both the ground and excited states of the system, with a particular focus on  the QFI near the EPs \cite{NingboQFI2023,LeeCh2021}.

This paper is organized as follows:  In Sec. II,  we  derive the transcendental function whose zeros provide the complete spectrum, using the BOA. The EPs are identified through their  real number form in the $\mathcal{PT}$-unbroken phase.	The previously  often used AA  is  further corrected analytically.	In Sec. III, we examine the atomic  dynamics and the VRS in the emission spectrum using several approaches. The novel characteristics of the VRS are explored and discussed in terms of analytical eigenstates. In Sec. IV,  we calculate the QFI in the PTQRM and its Hermitian counterpart, as well as for the  NHTLS, and perform the comparison of the QFI values near the EPs. Finally, a brief summary is provided in Sec. V.	The appendix presents the determinant of {the symmetric correction to the AA (the CAA)} and the selection rules for the eigenstates. 	

\section{Analytical solution}

The non-Hermitian qubit in  a single-mode cavity can be described by the
following Hamiltonian
\begin{equation}
H=-\frac{\Delta }{2}\sigma _{x}+\frac{i\epsilon }{2}\sigma
_{z}+\omega a^{\dag }a+g\left( a^{\dag }+a\right) \sigma _{z},
\label{Hamiltonian}
\end{equation}%
where the first two terms represent a non-Hermitian qubit with energy
splitting $\Delta $ and an imaginary bias $i\epsilon $, $\sigma _{x,z}$ are
the Pauli matrices, $a^{\dag }$ and $a$ denote the creation and annihilation
operators with cavity frequency $\omega $, and $g$ represents the
qubit-cavity coupling strength. Its Hermitian counterpart is just the asymmetric QRM{ \cite{Braak2011,qhchen2012,zhang2013,li2015,ashhab2020,mangazeev2021,li2021,xie2022}}.
For $\epsilon = 0$,  Hamiltonian (\ref{Hamiltonian}) exhibits $\mathbb{Z}_{2}$
symmetry [43,44]. The conserved parity operator is defined as $%
\hat{P} = -\sigma_{x}\exp(i\pi a^{\dag}a)$, with eigenvalues ${\pm 1}$. This
symmetry allows  Hamiltonian (\ref{Hamiltonian}) to be diagonalized into two blocks, each corresponding to a different parity subspace. Due to this $\mathbb{Z}%
_{2}$ symmetry, eigenvalues from different symmetry parity are allowed to
cross.

When a pure imaginary bias $\epsilon \neq 0$ is introduced, the parity
operator {$\hat{P}$ }is no longer conserved. Combined with the standard
time-reversal operator $T$, which takes the complex conjugate, and noting
that $\mathcal{T}^{\dagger}\hat{x}\mathcal{T} = \hat{x}$ and $\mathcal{T}%
^{\dagger}\hat{p}\mathcal{T} = -\hat{p}$, where $\hat{x}$ is the
displacement operator and $\hat{p}$ is the momentum operator, it can be
verified that $\left[H, \mathcal{PT}\right] = 0$. Therefore, the
non-Hermitian Hamiltonian (~\ref{Hamiltonian}) is indeed $\mathcal{PT}$%
-symmetric, and is thus termed the PTQRM.
The PTQRM can be implemented in a practical  system as proposed in Ref.~\cite{LiPT2023}.  After performing a rotation around the y axis in the atomic sector,  Hamiltonian (~\ref{Hamiltonian}) transforms into their Eq. (1). Consider a passive $\mathcal{PT}$-symmetric qubit with the Hamiltonian $H_q^p = -\frac{\Delta}{2} \sigma_x + i \epsilon \sigma_+ \sigma_-$, which differs from the standard $\mathcal{PT}$-symmetric qubit [i.e., first two terms in  (~\ref{Hamiltonian})] by a constant shift of $i\epsilon / 2$, but retains all the essential features of standard $\mathcal{PT}$ symmetry, including EPs. Incorporating a cavity into this passive qubit leads to the QRM under passive $\mathcal{PT}$ symmetry, known as the passive PTQRM.  Using the Lindblad master-equation(LME) approach, Lu et al. demonstrate that the effective Hamiltonian governing the dynamics is synonymous with the passive PTQRM in the rotating frame, and is equivalent  to the standard PTQRM Hamiltonian ~\cite{LiPT2023}. These authors have also proposed a potential circuit QED realization in which the transmon qubit circuit is embedded in a cavity.

{The effective non-Hermitian Hamiltonian typically emerges within the framework of quantum trajectories and postselection.	Reference [45] offers a detailed explanation of when quantum jumps can be neglected and clarifies the relationship between the master equation and the non-Hermitian approach.	Discarding quantum jumps is as phenomenological as introducing a complex energy shift.	In the absence of quantum jumps, the system evolves according to the effective non-Hermitian Hamiltonian. Repeating experiments and discarding trajectories have been experimentally demonstrated in Ref. ~\cite{naghiloo2019}, where an effective non-Hermitian Hamiltonian and the EPs for quantum systems were reported.		This suggests that experimentalists have already considered the effects of postselection in practical experiments.}  

In this section, we present the exact solution and several analytical
approximations to the PTQRM.

\subsection{Exact solutions by the Bogoliubov operator approach}

By two Bogoliubov transformations
\begin{equation}
A_{\pm }=a\pm g/\omega,  \label{shift}
\end{equation}%
Hamiltonian (\ref{Hamiltonian}) can be transformed into the following
matrix form with units $\hbar =\omega =1$:
\begin{equation}
H=\left(
\begin{array}{cc}
A_{+}^{\dagger }A_{+}+i\frac{\epsilon }{2}-g^{2} & -\Delta /2 \\
-\Delta /2 & A_{-}^{\dagger }A_{-}-i\frac{\epsilon }{2}-g^{2}%
\end{array}%
\right).  \label{matrix}
\end{equation}%
The wavefunction can be expressed as a series expansions in terms of the $%
A_{+}$ operator
\begin{equation}
\left\vert A_{+}\right\rangle =\left( \
\begin{array}{l}
\sum_{n=0}^{\infty }\sqrt{n!}e_{n}\left\vert n\right\rangle _{A_{+}} \\
\sum_{n=0}^{\infty }\sqrt{n!}f_{n}\left\vert n\right\rangle _{A_{+}}%
\end{array}%
\right) ,  \label{wave1}
\end{equation}%
where $e_{n}$ and $f_{n}$ are the expansion coefficients. It can  also be  expressed in terms of the $A_{-}$ operator.
\begin{equation}
\left\vert A_{-}\right\rangle =\left( \
\begin{array}{l}
\sum_{n=0}^{\infty }\left( -1\right) ^{n}\sqrt{n!}c_{n}\left\vert
n\right\rangle _{A_{-}} \\
\sum_{n=0}^{\infty }\left( -1\right) ^{n}\sqrt{n!}d_{n}\left\vert
n\right\rangle _{A_{-}}%
\end{array}%
\right) ,  \label{wave2}
\end{equation}%
with two coefficients, $c_{n}$ and $d_{n}$. The states $\left\vert
n\right\rangle_{A{+}}$ and $\left\vert n\right\rangle_{A{-}}$ are known as
extended coherent states, possessing the following properties:

\begin{eqnarray}
|n\rangle _{A_{\pm }} &=&\frac{\left( A_{\pm }^{\dagger }\right) ^{n}}{\sqrt{%
n!}}|0\rangle _{A_{\pm }}=\frac{\left( a^{\dagger }\pm g\right) ^{n}}{\sqrt{%
n!}}|0\rangle _{A_{\pm }},  \label{extended} \\
|0\rangle _{A_{\pm }} &=&e^{-(1/2)g^{2}\mp ga^{\dagger }}|0\rangle _{a},
\notag
\end{eqnarray}%
where the vacuum state $|0{\rangle }_{A_{\pm }}$ in Bogoliubov operators $%
A_{\pm }$ is well defined as the eigenstate of the original photon
annihilation operator $a$, and is known as a pure coherent state \cite%
{Glauber1963}.

By the    Schr\"{o}dinger equation, we get the linear relation for
two coefficients $e_{m}\;$and $f_{m}\;$of Eq. (\ref{wave1}) with the same
index $m$ as
\begin{equation}
e_{m}=\frac{\Delta }{2\left( m-g^{2}+\frac{i\epsilon }{2}-E\right) }f_{m},
\label{cor1}
\end{equation}%
and the coefficient $f_{m}$ can be defined recursively %\begin{widetext}
\begin{eqnarray}
\left( m+1\right) f_{m+1} &=&\frac{1}{2g}\left( m+3g^{2}-\frac{i\epsilon }{2}%
-E\right.  \notag  \label{fm} \\
&&\left. -\frac{\Delta ^{2}}{4\left( m-g^{2}+\frac{i\epsilon }{2}-E\right) }%
\right) f_{m}-f_{m-1},  \notag
\end{eqnarray}%
%
%\end{widetext}
with\ $f_{0}=1$. Similarly, the two coefficients $c_{m}\;$and $d_{m}$ of Eq.
(\ref{wave2}) satisfy
\begin{equation}
d_{m}=\frac{\Delta }{2\left( m-g^{2}-\frac{i\epsilon }{2}-E\right) }c_{m},
\label{cor2}
\end{equation}%
and the recursive relation is given by %\begin{widetext}
\begin{eqnarray}
\left( m+1\right) c_{m+1} &=&\frac{1}{2g}\left( m+3g^{2}+\frac{i\epsilon }{2}%
-E\right.  \notag  \label{cm} \\
&&\left. -\frac{\Delta ^{2}}{4\left( m-g^{2}-\frac{i\epsilon }{2}-E\right) }%
\right) c_{m}-c_{m-1},  \notag
\end{eqnarray}%
with $c_{0}=1$.

\begin{figure}[tbph]
\includegraphics[width=\linewidth]{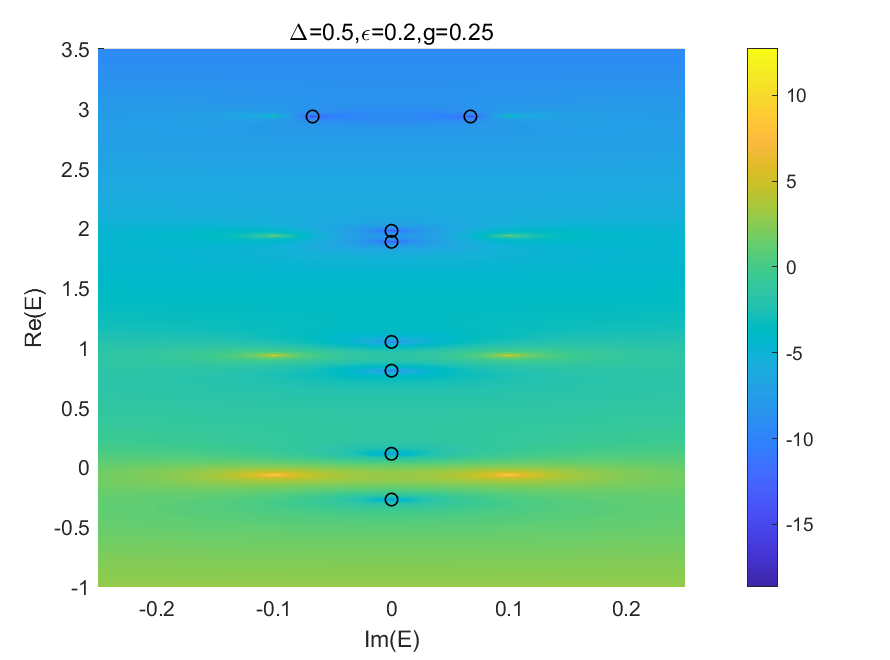}
\caption{ln$\lvert G\rvert^{2}$, where G is calculated from  $G$ function (\ref{Gge}), in the complex $E$ plane. Open circles mark the zeros of both the real and imaginary parts of the $G$ function, which coincide exactly with the eigenvalues from the ED.	 The  parameters are $\Delta =0.5$,  $\epsilon =0.2$, and $g=0.25$. }
\label{G_general}
\end{figure}

If both wavefunctions (\ref{wave1}) and (\ref{wave2}) are true
eigenfunctions for a nondegenerate eigenstate with eigenvalue $E$, they
should differ only by a constant ratio, i.e. $\left\vert A_{+}\right\rangle
=z\left\vert A_{-}\right\rangle $. Projecting both sides onto the original
vacuum state $\left\vert 0\right\rangle $, using $\sqrt{n!}{\langle }0|n{%
\rangle }_{A_{+}}=(-1)^{n}\sqrt{n!}{\langle }0|n{\rangle }%
_{A_{-}}=e^{-g^{2}/2}g^{n}$ and eliminating $z$ gives

\begin{equation}
\sum_{n=0}^{\infty }e_{n}g^{n}\sum_{n=0}^{\infty
}d_{n}g^{n}=\sum_{n=0}^{\infty }f_{n}g^{n}\sum_{n=0}^{\infty }c_{n}g^{n}.
\label{Gg}
\end{equation}%

{Thus, the $G$ function can be defined as}

{\begin{equation}
G=\sum_{n=0}^{\infty }e_{n}g^{n}\sum_{n=0}^{\infty
}d_{n}g^{n}-\sum_{n=0}^{\infty }f_{n}g^{n}\sum_{n=0}^{\infty }c_{n}g^{n}.
\label{G}
\end{equation}%
}

{By substituting Eqs. (\ref{cor1}) and (\ref{cor2}), one can  arrive at}

\begin{eqnarray}
G &=&\left( \frac{\Delta }{2}\right) ^{2}\left[ \sum_{n=0}^{\infty }\frac{%
f_{n}}{n-g^{2}+\frac{i\epsilon }{2}-E}g^{n}\right]  \notag \\
&&\times \left[ \sum_{n=0}^{\infty }\frac{c_{n}}{n-g^{2}-\frac{i\epsilon }{2}%
-E}g^{n}\right]  \notag \\
&&-\sum_{n=0}^{\infty }f_{n}g^{n}\sum_{n=0}^{\infty }c_{n}g^{n}.  \label{Gge}
\end{eqnarray}
  It includes both real and imaginary components. The zeros of this $G$ function provide both the real and imaginary components of the eigenvalues. As illustrated  in Fig.~\ref{G_general}, the eigenvalues obtained from the  $G$ function  (\ref{Gge}) show excellent agreement with those from exact diagonalization (ED).

\begin{figure}[tbph]
\centering
\includegraphics[width=\linewidth]{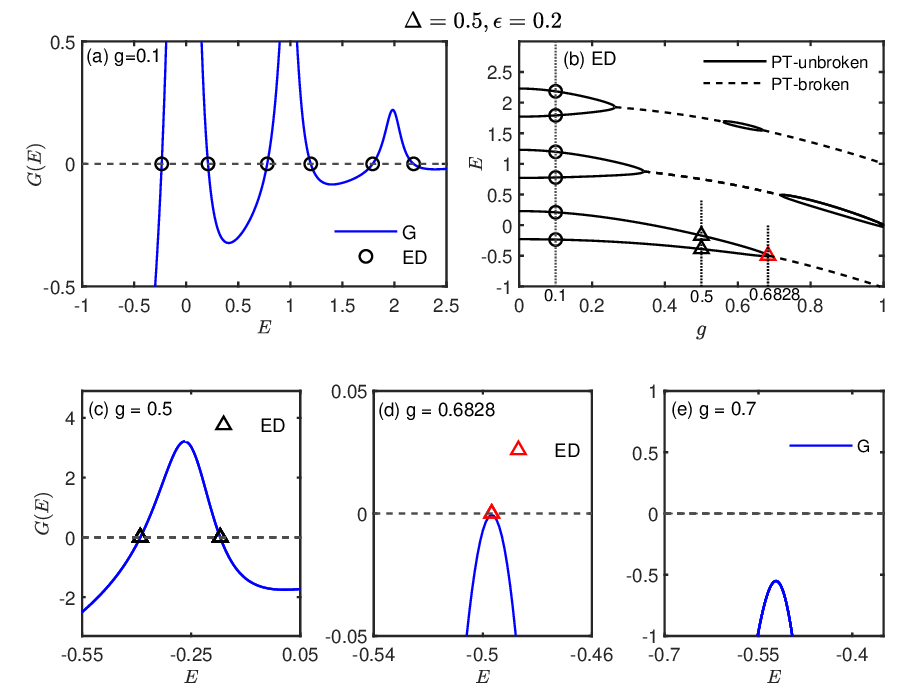}
\caption{(a) $G$ curves from Eq. (~\ref{Gfunc}) (blue line) at $g=0.1$ in the real eigenvalue regime, with circles representing ED results. (b) The real part of  eigenvalues as a function of g from the ED, with the dashed  lines  marking the real  part of eigenvalues in the $\mathcal{P}\mathcal{T}$-broken region. Crossing values between the  vertical dotted lines at $g=0.1,0.5$ and $0.6828$ and real eigenvalues (solid line) in the $\mathcal{P}\mathcal{T}$-unbroken region are confirmed in the $G$ functions in (a) and the lower panel.   $G$ curves from Eq. (~\ref{Gfunc})  at $g=0.5,0.6828$ and $0.7$  are shown in  (c), (d), and (e) with blue lines, respectively.  The  parameters are$\Delta=0.5$ and$\epsilon = 0.2$. }
\label{G_fig}
\end{figure}

Notably, in the $\mathcal{PT}$-unbroken phase, it can be shown that
\begin{eqnarray*}
e_{m} &=&d_{m}^{\ast }, \\
f_{m} &=&c_{m}^{\ast }.
\end{eqnarray*}%
By following the same procedure, one can derive the real $G$ function in the form
\begin{equation}
G_{NH}^{(s)}=\left\vert \sum_{n=0}^{\infty }e_{n}g^{n}\right\vert ^{2}-
\left\vert \sum_{n=0}^{\infty }f_{n}g^{n}\right\vert ^{2}.  \label{Gfunc}
\end{equation}%
Although the coefficients $e_{n}$ and $f_{n}$   are complex, both terms in Eq. (\ref{Gfunc}) represent squares of the modulus, making the $G$ function real. Its zeros can determine all real eigenvalues of this model within the $\mathcal{PT}$-unbroken region.

As shown in Fig.~\ref{G_fig}, all real  eigenvalues  (solid line) can be obtained by zeros of this real $G$ function.  EPs occur when both the $G$ function in Eq. (\ref{Gfunc}) and its first-order derivative with respect to E are simultaneously zero, as illustrated  in Fig.~\ref{G_fig} (d). The zero of the first-order derivative identifies  the EP where two real eigenvalues merge into one. Thus, using the  real $G$ function in Eq. (\ref{Gfunc}), we can accurately pinpoint  the position of the EPs and show that both eigenstates and eigenvalues coalesce at these points. It should be noted that, while we benefit from using the real $G$ function  (\ref{Gfunc})  to determine the EPs, it cannot be applied to  $\mathcal{PT}$-broken region, marked by the dotted line in Fig.~\ref{G_fig} (b) and the nonsolution in Fig.~\ref{G_fig} (e).

Finally, from Eqs. (\ref{cor1}) and (\ref{cor2}), it is clear  that the denominator cannot vanish for any $E$ due to the presence of the imaginary bias $\epsilon$. As a result, the pole structure of the $G$ functions does not exist, in sharp contrast to its Hermitian counterpart [23,26]. In other words, all eigenvalues can be given by the  zeros of the general $G$ function (\ref{Gge}), and all real eigenvalues in the $\mathcal{PT}$ -unbroken phase can be obtained by zeros of the real $G$ function Eq.(\ref{Gfunc}), even at the EPs, in the PTQRM. Note that the doubly degenerate solutions in its Hermitian counterpart cannot be given by the corresponding  $G$ function [25,28].

\subsection{Adiabatic approximation method}

A previous study of the PTQRM \cite{LiPT2023} primarily discussed the AA. In this section, we present a simpler alternative to the AA method.  Our goal is to establish a practical
framework for the further correction and systematic improvement of the AA
results.

Based on the matrix form of Eq. (\ref{matrix}), the general wavefunction can
also be expanded in terms of the two new operators $A_{+}$ and $A_{-}$
simultaneously, as
\begin{equation}
\left\vert \Psi \right\rangle =\left(
\begin{array}{l}
\;\sum_{n=0}u_{n}\left\vert n\right\rangle _{A_{+}} \\
\sum_{n=0}(-1)^{n}v_{n}\left\vert n\right\rangle _{A_{-}},%
\end{array}%
\right) .  \label{wavefunction}
\end{equation}%
This ansatz for the wavefunction is based on the extended photonic coherent
states in Eq. (\ref{extended}), which rely on various polaron-like
transformations or shifted operators.

By the Schr\"{o}dinger equation, we have
\begin{eqnarray}
\left( m-g^{2}+i\frac{\epsilon }{2}\right) u_{m}-\sum_{n=0\ }D_{mn}v_{n}
&=&Eu_{m},  \label{exact1} \\
\left( m-g^{2}-i\frac{\epsilon }{2}\right) v_{m}-\sum_{n=0}D_{mn}u_{n}
&=&Ev_{m},  \label{exact2}
\end{eqnarray}%
where
\begin{equation*}
D_{mn}=\frac{\Delta }{2}\left( -1\right) ^{m}{}_{A_{-}}\left\langle
m\right\vert \left\vert n\right\rangle _{A_{+}},
\end{equation*}%
\begin{equation*}
_{A_{-}}\left\langle m\right\vert \left\vert n\right\rangle
_{A_{+}}=(2g)^{n-m}\exp (-2g^{2})\sqrt{\frac{m!}{n!}}L_{m}^{n-m}\left(
4g^{2}\right)
\end{equation*}%
For $n\geq m$, $L_{m}^{n-m}(x)$ represents a Laguerre polynomial, with  $%
D_{mn}=D_{nm}$. The term $_{{A}_{{-}}}\left\langle m\right\vert \left\vert
n\right\rangle_ {A_{+}}$ describes the overlap between the extended coherent
states $\left\vert m\right\rangle _{{A}_{{+}}}$ and $\left\vert
n\right\rangle _{{A}_{{-}}}$. We refer to both $\left\vert m\right\rangle _{{%
A}_{{+}}}$and $\left\vert m\right\rangle _{A_{-}}$ as the $m$th manifold in the space of the Bogoliubov operators below.

{For the truly exact solution, the summation in the second term of Eqs. (\ref{exact1}) and (\ref{exact2}) must be infinite.	Physically, tunneling between all the $m$th manifolds must be considered. 	The overlap $D_{mm}$ within the same $m$th manifold is  larger than the overlaps $D_{mn}$ between different manifolds ($n \neq m$).	Based on these observations, a simple preliminary analytical approach is to retain only the transitions between states within the same manifold.This is actually the adiabatic approximation introduced by Irish et al. for the Hermitian QRM~\cite{tw2005} and later extended to the two-qubit QRM~\cite{agarwal2012}. Further corrections can be made by including the transitions between different manifolds.	Therefore,} within the framework of Eqs. (\ref{exact1}) and (\ref{exact2}), analytical approximations can be systematically performed.	First, as a zero-order approximation, we omit the off-diagonal terms and retain only the terms within the same $m$th manifold，
\begin{equation}
\left\vert \Psi \right\rangle ^{(AA)}=\left(
\begin{array}{l}
\;u_{m}\left\vert m\right\rangle _{A+} \\
(-1)^{m}v_{m}\left\vert m\right\rangle _{A_{-}}%
\end{array}%
\right) ,  \label{adb}
\end{equation}%
and have
\begin{eqnarray}
\left( m-g^{2}+i\frac{\epsilon }{2}-E\right) u_{m}-\;D_{m,m}v_{m} &=&0,
\label{EX_AD1} \\
-D_{m,m}u_{m}+\left( m-g^{2}-i\frac{\epsilon }{2}-E\right) v_{m} &=&0.
\label{EX_AD2}
\end{eqnarray}
The nonzero coefficients $%
u_{m}$ and $v_{m}$ yield the following equation:
\begin{equation*}
\left( m-g^{2}+i\frac{\epsilon }{2}-E\right) \left( m-g^{2}-i\frac{\epsilon
}{2}-E\right) -\left( D_{m,m}\right) ^{2}=0.
\end{equation*}%
The $m$-th pair of eigenvalues is then given by
\begin{equation}
E_{\pm }=m-g^{2}\mp \frac{1}{2}\sqrt{4\left( D_{m,m}\right) ^{2}-\epsilon
^{2}},  \label{EZero}
\end{equation}%
where $m=0,1,2...$, and the corresponding $m$-th pair of eigenstates is
\begin{equation}
\left\vert m\right\rangle _{\pm }\varpropto \left(
\begin{array}{l}
\;\;\;\;\;\left( -1\right) ^{m}D_{m,m}\left\vert m\right\rangle _{A_{+}} \\
\left( m-g^{2}-i\frac{\epsilon }{2}-E_{\pm }\right) \left\vert
m\right\rangle _{A_{-}}%
\end{array}%
\right) .  \label{psim}
\end{equation}

Note that in Eq. (\ref{EZero}), complex eigenvalues appear if $\epsilon
>2D_{m,m}$,
\begin{equation}
E_{\pm }=m-g^{2}\mp \frac{i}{2}\sqrt{\epsilon ^{2}-4\left( D_{m,m}\right)
^{2}},
\end{equation}%
corresponding to the $\mathcal{PT}$-broken phase. Substituting $E_{+}$ into either Eq. (\ref{EX_AD1}) or Eq. (\ref{EX_AD2}) yields
\begin{eqnarray*}
\left( m-g^{2}+i\frac{\epsilon }{2}-E_{+}\right) u_{m}-\;D_{m,m}v_{m} &=&0,
\\
\frac{1}{2{D_{m,m}} }i\left( \sqrt{\epsilon ^{2}-4\left( D_{m,m}\right) ^{2}}%
+\epsilon \right) u_{m} &=&v_{m}.
\end{eqnarray*}%
Setting $c_{m}=1$, we have
\begin{equation}
\left\vert \Psi \right\rangle ^{(AA)}=N\left(
\begin{array}{l}
\;\left\vert m\right\rangle _{A_{+}} \\
\frac{1}{2D_{m,m}}i\left( \sqrt{\epsilon ^{2}-4\left( D_{m,m}\right) ^{2}}%
+\epsilon \right) \left\vert m\right\rangle _{A-}%
\end{array}%
\right), \label{psim_n}
\end{equation}%
where
\begin{equation*}
N=\left[ 1+\frac{1}{4D_{m,m}}\left( \sqrt{\epsilon ^{2}-4\left(
D_{m,m}\right) ^{2}}+\epsilon \right) ^{2}\right] ^{-1/2}.
\end{equation*}%
It can be confirmed that  wavefunction (\ref{psim_n}) lacks $\mathcal{PT}$%
 symmetry, even though the Hamiltonian still commutes with the $\mathcal{P}%
\mathcal{T}$ operator. This behavior contrasts with the case where the
system is in the $\mathcal{PT}$-unbroken phase. As an approximate
solution, it can be verified that $\left\vert m\right\rangle _{\pm }$ must
coincide at the EPs where {$\epsilon =2D_{m,m}$}, as given in { Eq. (19)}.

Note that the zero-order approximation here is identical to the AA
discussed by Lu et al. \cite{LiPT2023}. The transitions between different
manifolds $\left\vert m\right\rangle _{{A{\pm }}}$ are omitted in this
approximation of the wavefunction (\ref{adb}), which is commonly referred to
as the adiabatic approximation.

The AA results for the eigenvalues, shown in Fig.~\ref{spectrum_CAA} for $\protect\epsilon = 0.1$  and $\protect\epsilon = 0.5$, basically align with the ED results but still deviate in both the $\mathcal{PT}$-unbroken and $
\mathcal{P}\mathcal{T}$-broken region, except in the deep strong-coupling regime. There is still room for improving the AA.	
The key assumption of the AA is that only transitions within the same manifolds are considered, while transitions between different manifolds are neglected \cite{GAA2021}, {rendering the AA approach trivial}. To obtain more accurate results, additional transitions must be included.	Corrections to the AA method are discussed in the next section.

\subsection{Corrections to the adiabatic approximation}

The AA method can be incrementally improved by symmetrically including additional off-diagonal elements in the current formalism.	The symmetric correction to the AA, referred to as CAA, is made by selecting three sets of coefficients: $\left( u_{m-1},v_{m-1}\right)$ , $\left(u_{m},v_{m}\right)$  and $\left( u_{m+1},v_{m+1}\right) $, which represent symmetric transitions between the two nearest-neighboring manifolds and within the same manifold.	
$m\leftrightarrow m,m\pm 1$. In contrast, in the AA,  the $(m\pm1)$-th manifold is not involved in the $m$-th pair of eigenvalues.

\begin{figure}[tbph]
\includegraphics[width=\linewidth]{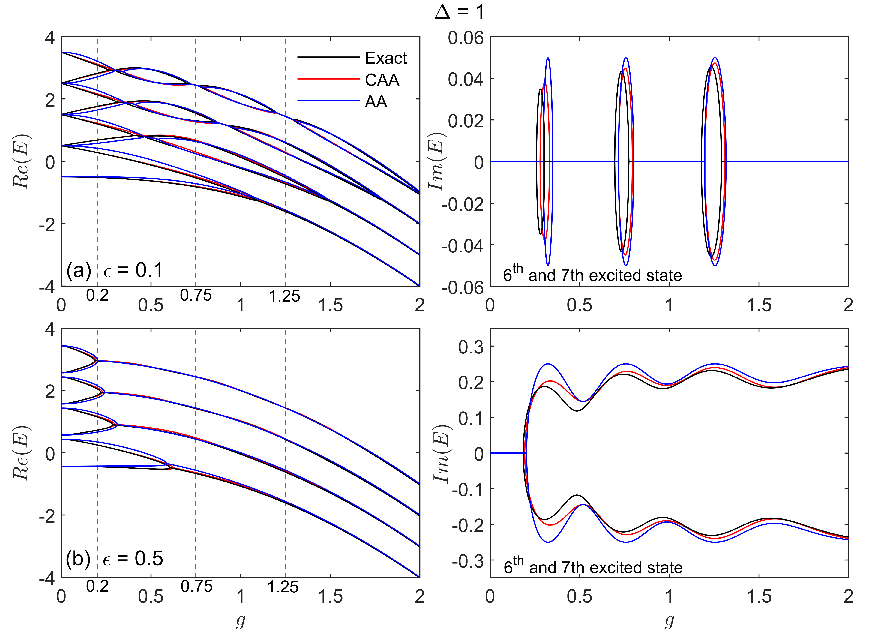}
\caption{ The real part (left) of the first four pairs and the imaginary part (right) of only the fourth pair of eigenvalues (for clarity) as a function of coupling strength $g$ at resonance $\Delta = 1$ for $\epsilon = 0.1$ (upper panel) and $\epsilon = 0.5$ (lower panel).	The black lines represent the ED results, the blue dots show the AA results, and the red dots indicate the CAA results.	The eigenvalues and corresponding eigenstates along the dashed vertical line will be used to calculate the dynamics shown in Fig. \ref{Dynamic_ultra}, \ref{Dynamic_near_deep} and \ref{Dynamic_deep}.}
\label{spectrum_CAA}
\end{figure}
According to Eq. (\ref{wavefunction}), the eigenstate corresponding to the $m$-th pair eigenvalues is selected as %\begin{widetext}
\begin{eqnarray}
&&\left\vert \Psi _{CAA}\right\rangle = \\
&&\left(
\begin{array}{l}
\; u_{m-1}\left\vert m-1\right\rangle _{A_{+}}+u_{m}\left\vert
m\right\rangle _{A_{+}} + u_{m+1}\left\vert m+1\right\rangle _{A_{+}} \\
\mp v_{m-1}\left\vert m-1\right\rangle _{A_{-}}\pm v_{m}\left\vert
m\right\rangle_{A_{-}}\mp v_{m+1}\left\vert m+1\right\rangle _{A_{-}}%
\end{array}%
\right) ,  \notag  \label{SOA}
\end{eqnarray}
where the first $\mp$ in the second line corresponds to odd and even $m$.
The $m$-th pair of eigenvalues can be obtained by selecting the determinant
in a $6\times6$ block form with $ n=m,m\pm 1 $ in Eqs. (\ref{exact1}) and (\ref{exact2}), as presented in the Appendix. This determinant leads
to a polynomial equation of degree $6$. An explicit analytical solution
cannot be obtained due to the Abel-Ruffini theorem, which asserts  that no
general solution in radicals exists for polynomial equations of degree $5$
or higher with arbitrary coefficients \cite{fraleigh2003}. There are six
groups of roots, but only two are required. The CAA results should match the
AA results from Eq. (\ref{EZero}). This allows us to unambiguously identify the $m$-th pair of eigenvalues in the CAA from the full set of solutions. A simple selection rule is outlined in the Appendix.

Figure~\ref{spectrum_CAA} shows that the CAA results for the eigenvalues  align well with
the exact ones in the entire coupling regimes. In particular,  near the EPs, the CAA results show a marked improvement over the AA. However, the key improvement is not in the accuracy of the eigenvalues, but in the form of the eigenstates, which will be clearly illustrated in the dynamics.	

Further corrections, such as the inclusion of five sets of coefficients, $\left( u_{m\pm i}, v_{m\pm i}\right), i = 0, 1, 2$, can also be implemented in a similar manner.	In principle, these corrections can be applied systematically, step by step, ultimately converging to the exact solutions outlined in Sec. II A.	However, if a qualitative understanding of the physics is achieved within the CAA, additional corrections may be unnecessary in the analytical sense, as shown in the next section, where the CAA will also be used to calculate the qubit dynamics and study the VRS of the PTQRM.	

\section{Dynamics and vacuum Rabi splitting}
One of the notable features in atom-cavity coupling system is  the  VRS, as described in Refs.\cite{agarwal1984,thompson1992,boca2004}. The atom is typically pumped from its ground state to the excited state, and then  decays to the ground state of the whole system. The resulting emission spectrum displays two peaks of equal height at resonance, characteristic of  VRS in the Jaynes-Cummings model~\cite{sanchez1983}.

In this section, we calculate the time evolution of the qubit population difference, $\left\langle \sigma _{z}\right\rangle $, in the PTQRM using three methods:  the CAA, the AA, and the exact solution, either ED or through  the exact eigensolution obtained via the BOA.
We use $\left|\psi_n^R\right\rangle$ and $\left|\psi_n^L\right\rangle$ to denote the right and left eigenstates, the eigenstates of  $H$ and $H^{\dagger}$, respectively. The normalization conditions are  $\left\langle\psi_n^R \mid \psi_n^R\right\rangle=1$ and $\left\langle\psi_m^L \mid \psi_n^R\right\rangle=$ $\delta_{m, n}$, while the right and left eigenstates with different eigenvalues are nonorthogonal. That is $\left\langle\psi_m^{R(L)}\mid \psi_n^{R(L)}\right\rangle \neq 0$ for $m \neq n$, due to the non-Hermitian nature ~\cite{Brody2014}. The time-dependent  wave function of the PTQRM, starting from  $|\Psi(0)\rangle$,  can be expressed as
\begin{equation}
|\Psi(t)\rangle=e^{-i H t}|\Psi(0)\rangle=\sum_m e^{-i t E_m}\left|\psi_m^{R}\right\rangle\left\langle\psi_m^{L} \mid \Psi(0)\right\rangle,
\end{equation}
$\left\langle \sigma _{z}\right\rangle $ is thus given by {
\begin{equation}
\left\langle \sigma _{z}(t)\right\rangle =\frac{\left\langle \Psi
(t)\left\vert \sigma _{z}\right\vert \Psi (t)\right\rangle }{\left\Vert \Psi
(t)\right\Vert ^{2}},  \label{sigmaz}
\end{equation}%
where $\left\Vert \Psi (t)\right\Vert =\sqrt{\langle \Psi (t)\mid \Psi
(t)\rangle }$} is the norm of the wavefunction, which is necessary  in non-Hermitian systems.  In general non-Hermitian Hamiltonians describe the
dynamics of physical systems that are not conservative.  The quantity $\langle \sigma_z(t) \rangle$ is related to the qubit population in the literature through the expression $(1 + \langle \sigma_z(t) \rangle) / 2$. This provides a more direct measure of the upper state the qubit.	

To analyze the dynamic behavior in more detail, a Fourier spectral analysis of the atomic population's time evolution is performed.	 The Fourier transform, a key  technique for frequency decomposition, is defined as follows:
\[
\mathcal{F}(\nu )=\int_{0}^{\infty }<\sigma_{z}(t)>e^{-i2\pi \nu t}dt,
\]%
where $\mathcal{F}(\nu)$ represents the emission spectrum, and $F(t)$ denotes the observed time evolution.	

\begin{figure}[tbph]
\includegraphics[width=\linewidth]{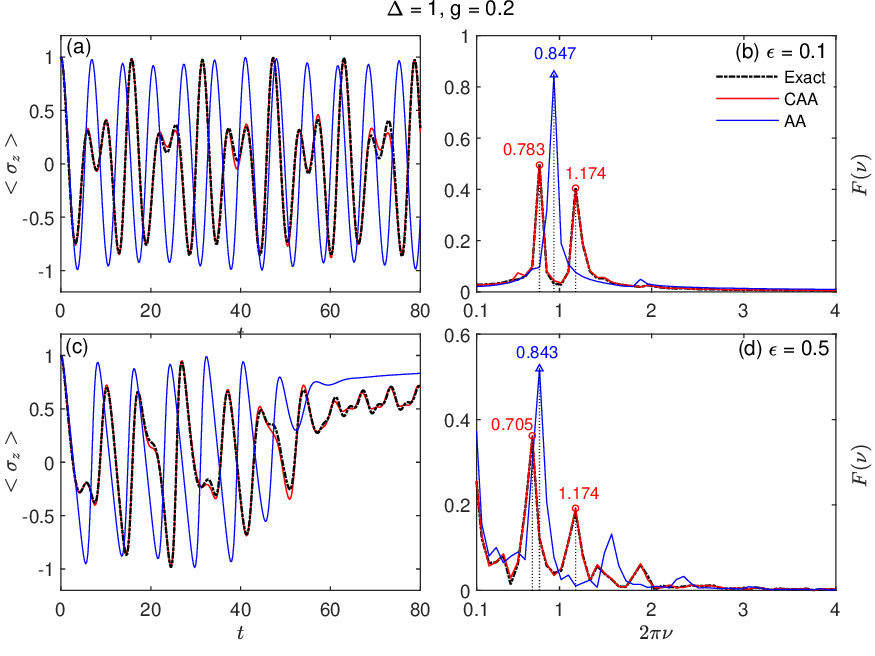}
\caption{ Time evolution of  $<\sigma_{z}>$ (left) and the corresponding emission spectrum (right) at  resonance ($\Delta = 1$) for the ultrastrong coupling ($g=0.2$) with $\epsilon = 0.1$ (upper panel) and $\epsilon = 0.5$ (lower panel). The peak positions are also shown in the left-hand panel.  }
\label{Dynamic_ultra}
\end{figure}

\begin{figure}[tbph]
\includegraphics[width=\linewidth]{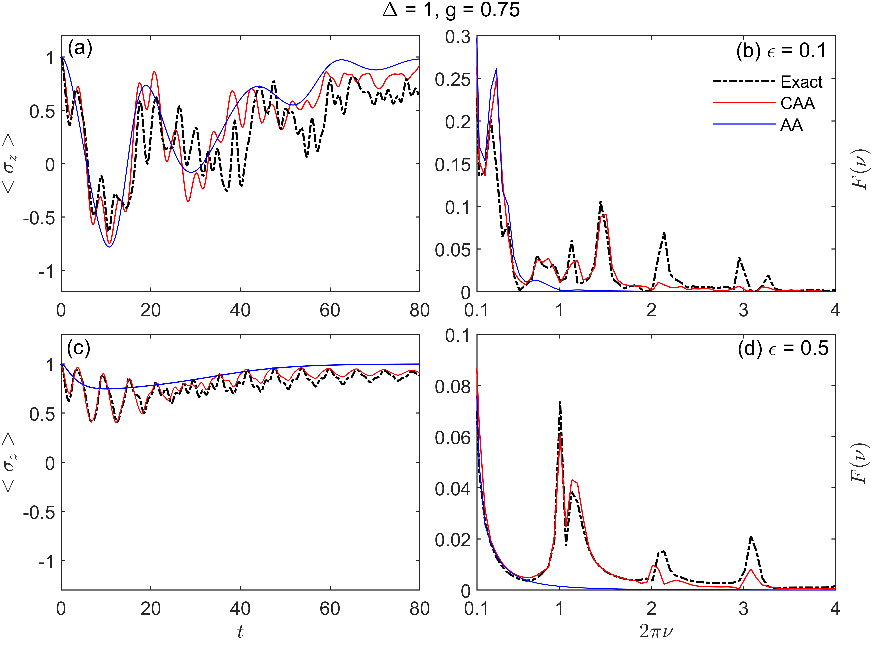}
\caption{ Time evolution of  $<\sigma_{z}>$ (left) and the corresponding emission spectrum (right) at  resonance ($\Delta = 1$) for near deep strong coupling ($g=0.75$) with $\epsilon = 0.1$ (upper panel) and $\epsilon = 0.5$ (lower panel).}
\label{Dynamic_near_deep}
\end{figure}

\begin{figure}[tbph]
\includegraphics[width=\linewidth]{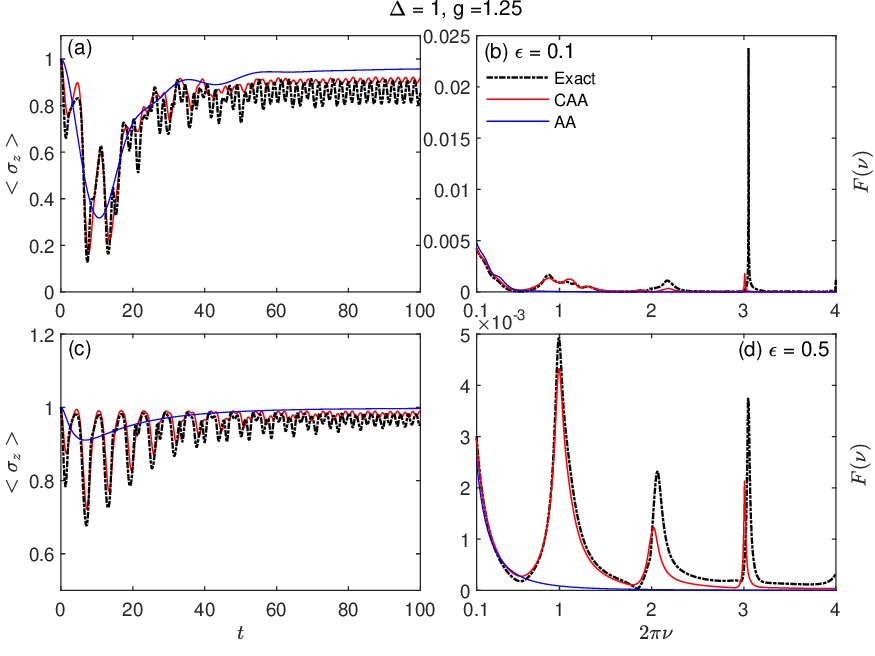}
\caption{ Time evolution of  $<\sigma_{z}>$ (left) and the corresponding emission spectrum (right) at  resonance ($\Delta = 1$) for the deep strong coupling ($g=1.25$) with $\epsilon = 0.1$ (upper panel) and $\epsilon = 0.5$ (lower panel).}
\label{Dynamic_deep}
\end{figure}

The initial state is chosen as the atomic upper level and the photon vacuum state  in the PTQRM, similar to the conventional study of the VRS. The dynamical evolution at resonance ($\Delta = 1$) is calculated using the CAA for two imaginary biases, $\epsilon = 0.1$ and $\epsilon = 0.5$, and three typical coupling strengths: $g = 0.1$, $0.75$, and $1.25$. These coupling strengths range from the ultrastrong to deep-strong coupling regimes. The results (red line) for the dynamics are respectively shown in the left-panel of Figs.\ref{Dynamic_ultra}, ~\ref{Dynamic_near_deep}, and ~\ref{Dynamic_deep}, while the emission  spectrum of the atomic population evolution is shown in the right-panel of these three figures.	Results from the ED ( dashed black line) and AA (blue line) are also presented for comparison.	

The present initial state  has dominant components in the low-energy excited states of the system.	 In the early stages of evolution, it involves only the ground stat and the lower excited states, with progressively higher excited states participating as time progresses.	  Generally, EPs shift toward lower coupling strengths ($g$) as the excited state number increases,  as seen in the left panel of Fig.~\ref{spectrum_CAA}. As indicated by the vertical dashed line, the spectrum at $g = 0.2$ is partially in the $\mathcal{PT}$-unbroken phase for lower excited states, while the spectra at $g = 0.75$ and $g = 1.25$ are primarily in the $\mathcal{PT}$-broken phase for most lower excited states.	We now proceed with the three typical coupling regimes.

\textsl{Ultrastrong coupling regime ($g=0.2$).} As shown in Fig.~\ref{Dynamic_ultra}, the CAA results perfectly agree with the exact ones  in both atomic dynamics and the emission spectrum. Oscillations with two main frequencies  are observed in the PTQRM, similar to the  VRS in its Hermitian counterpart. This  demonstrates that $\mathcal{PT}$-symmetry is preserved in the local phase during the early stages of the time evolution.   In contrast, a oscillation with a single frequency is exhibited in the AA. Note that subharmonic peaks in the high frequency region of the emission spectrum result from the imperfect sinusoidal oscillations and do not indicate oscillations at new frequencies.

Interestingly,  the positions of the two peaks in the CAA coincide with the crossing points of the vertical dashed line and the CAA eigenvalues in the left-hand panel  of Fig.~\ref{spectrum_CAA}, corresponding to the energies for the first and second excited states relative to the ground state energy,  $E^{CAA}_1-E^{CAA}_{GS}$ and $E^{CAA}_2-E^{CAA}_{GS}$, respectively. In contrast, the position of the single peak in the AA reflects the difference between the first excited state energy ($E_{+}$) and the ground state energy ($E_{-}$), as given  in Eq.(\ref{EZero}), and is also shown in the left-hand panel of Fig.~\ref{spectrum_CAA}. This occurs because, in the CAA, the nearest-neighboring manifold is considered, allowing transitions from both the first and second excited states to the ground state.	In contrast, only the transition between the same manifold is taken into account in  the AA, so the second excited state is excluded.

\textsl{Near deep-strong coupling regime ($g=0.75$).-} As shown in the upper panel of Fig.~\ref{Dynamic_near_deep} for both biases, the population from both the CAA and exact solutions oscillates with more than two frequencies, indicating multiple VRS. This is similar to the multiple Rabi oscillations observed in the ultrastrong coupling regime of its Hermitian counterpart without biases~\cite{zhangVRS2013} and in systems with ultrastrong molecular vibrational coupling~\cite{george2016}. However, the frequency spectrum in the AA shows only one peak for $\epsilon = 0.1$, and the $<\sigma_z(t)>$ curve tracks only the envelope of the multiple oscillations observed in the exact solutions, erasing the details of the fast oscillations.	 No peak at finite frequency is observed in the AA for large bias $\epsilon = 0.5$.

Multiple peaks in the frequency spectrum arise in the CAA because higher $m$ pairs of eigenstates are increasingly involved as the coupling strength increases, covering more manifolds than the $m=0$ pair used in the ultrastrong coupling regime ($g = 0.2$). For example, the $m=1$ pair of eigenstates spans the $m=0,1,2$ manifolds, while the $m=0$ pair only spans the $m=0,1$ manifolds.

\textsl{Deep-strong coupling regime $g=1.25$.-} In this coupling regime, the atomic population difference never becomes negative, indicating that the atom predominantly stays in the upper level. The atomic population oscillates also with multiple frequencies. The CAA matches the exact solution, while the AA only captures the general envelope of the true dynamics, eliminating all the fast oscillations.	

In the three typical coupling regimes discussed above, it is evident that the AA does not accurately capture the true behavior of atomic population evolution in this non-Hermitian model.	The CAA results match the ED results in all cases. 	 Although the eigenvalue difference between the AA and CAA is relatively small, the essential differences in dynamics can still depend on the number of manifolds considered.  	In other words,  transitions within the same manifold are insufficient to describe  the dynamics; transitions between different manifolds must also be considered in the analysis.	Since the CAA provides an accurate and precise description of the dynamic behavior of the PTQRM in all cases, the additional consideration of transitions between nearest-neighboring manifolds is already sufficient  to capture the key physics of the model.

In this non-Hermitian $\mathcal{PT}$-symmetric system, the atom tends to return to its original state, accompanied by  an infinite number of photons~\cite{LiPT2023}. This feature is clearly visible  in Figs.~\ref{Dynamic_ultra}(c), ~\ref{Dynamic_near_deep}, and ~\ref{Dynamic_deep} during the final time window. For the small bias and  ultrastrong coupling regime, we calculated   the long-time evolution, continuing from the data in Fig.~\ref{Dynamic_ultra}(a), and  observed    similar behavior after a prolonged oscillation (not shown here) . This represents the final outcome,  likely due to the balance of gain and loss in the non-Hermitian system with  $\mathcal{PT}$ symmetry.	

\begin{figure}[tbph]
\includegraphics[width=\linewidth]{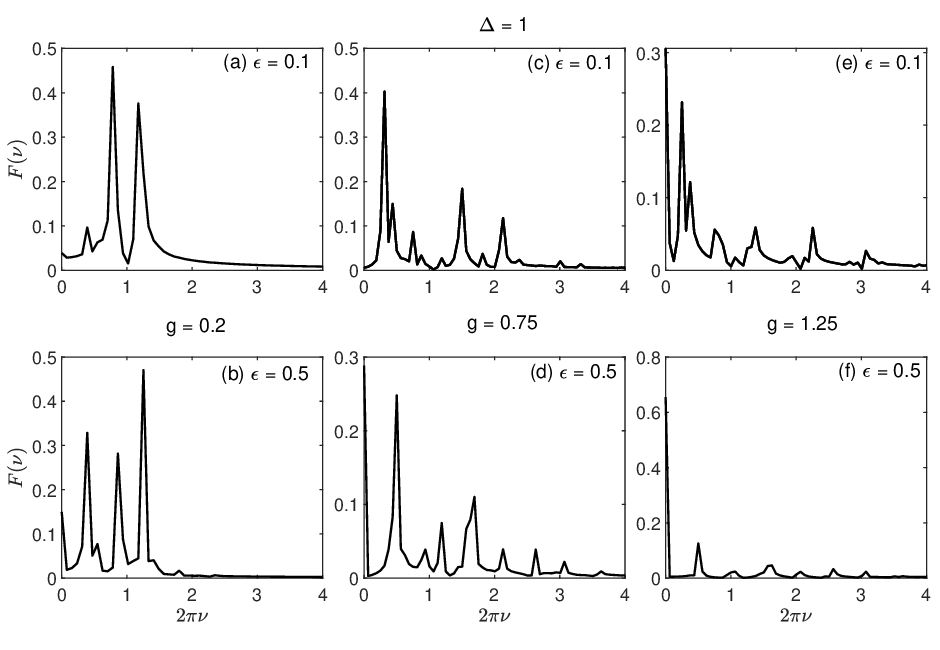}
\caption{The frequency spectrum of the Hermitian QRM at resonance ($\Delta = 1$) for  ultrastrong coupling ($g=0.2$), the near deep-strong coupling ($g=0.75$), and the deep-strong coupling ($g=1.25$), with $\epsilon = 0.1$ (upper panel) and $\epsilon = 0.5$ (lower panel).}
\label{Dynamic_HQRM}
\end{figure}

As the coupling strength increases, the VRS splits into multiple Rabi frequencies. The width of the VRS peaks increases with both the bias and the coupling strength, due to the  $\mathcal{PT}$-broken phases involved.  As shown in Fig. ~\ref{Dynamic_HQRM}  for its Hermitian counterpart with the  same real biases, the VRS exhibits sharp peaks. Multiple VRS also emerges with $\epsilon$, but the width remains unchanged.	The peak width  is typically determined by the decay rate. The imaginary bias, $i\epsilon$,  effectively plays the role of a decay rate. The pronounced  width of the VRS peaks resembles  that of the VRS in two-level atoms of an atom-cavity system dissipating in a bosonic environment ~\cite{orszagM2024}.	 This suggests that non-Hermiticity can be viewed as an effective description of an open system, assuming quantum jumps are disregarded [45].

\section{quantum Fisher information and enhanced quantum sensing}

The QFI provides the optimal accuracy limit for
parameter estimation and has significant applications in quantum precision
measurement. \cite%
{NingboQFI2023,QFI1994,QuantumMetrology2011,ying2024,ding2023}%
. In physical systems with critical properties, small changes in parameters
near a critical point can cause significant responses in the quantum state,
enabling the exploitation of this critical enhancement effect to improve
parameter estimation accuracy. Increasing the QFI to improve the estimation
accuracy of unknown parameters is a key issue in quantum metrology. Recent
theoretical and experimental studies have shown that near EPs unique to non-Hermitian systems, the system exhibits a strong response
to perturbations in parameters. Therefore, EPs can be used to achieve
high-precision sensing \cite{yu2020,liao2021}.

In quantum parameter estimation theory, the QFI
describes the minimum error in extracting an unknown parameter from a given
quantum state. The optimal measurement accuracy is governed by the quantum
Cram\'er-Rao bound ~\cite{QFI1994,sidhu2020,NingboQFI2023}. It can be
proven that the expressions for both the traditional QFI and the parameter
estimates of the quantum Cram\'{e}r-Rao bound still hold for
non-Hermitian Hamiltonians under stationary-states \cite%
{NingboQFI2023,yu2024,ren2024a}. Thus, for the PTQRM, the QFI for the pure states
is defined as {\
\begin{equation}
\mathcal{F}_{\lambda }(|\psi \rangle )=4\left( \left\langle \psi ^{\prime
}(\lambda )\mid \psi ^{\prime }(\lambda )\right\rangle -\left\vert
\left\langle \psi ^{\prime }(\lambda )\mid \psi (\lambda )\right\rangle
\right\vert ^{2}\right),  \label{QFI}
\end{equation}
where $\lambda $ represents the  experimental  parameter  to be estimated. In the present study, we select $\lambda =g$, $|\psi (g)\rangle $ is one eigenstate and $|\psi ^{\prime }(g)\rangle $ is its first-order derivative with respect to the coupling strength $g$. As mentioned previously, for a non-Hermitian Hamiltonian ($H^{\dag }\neq H$), the eigenstates are typically nonorthogonal. Under the stationary state condition, the normalization condition of the eigenstates holds, and it can be proven that the QFI still
satisfies Eq. (\ref{QFI}). We have verified  that $|\psi (g)\rangle $ can be selected either the right or the left eigenstates of the PTQRM, the QFI remains unchanged.

{Both the static ~\cite{ying2022c,montenegro2021} and dynamic ~\cite{garbe2020a,chu2021} approaches are commonly used  in  the critical quantum metrology protocol. In this work, we focus on measuring the QFI in equilibrium, a method that falls under the static approach. Figure~\ref{QCE} (a) illustrates  the difference in QFIs between the PTQRM and  the Hermitian QRM  with $\Delta = 0.5$ in the ground state as a function of coupling strength $g$ and $\epsilon$, calculated using either the wavefunction derived from the $G$ function technique or the ED. Throughout  the whole parameter regime plotted (up to the deep strong coupling $g=1.5$), the QFI of the PTQRM exhibits  a significant enhancement near the EPs compared to the Hermitian QRM. Figure~\ref{QCE} (c) shows the results for various bias parameters ($\epsilon = 0, 0.2, 0.4$).} The QFI in the PTQRM is significantly  higher than those in the Hermitian QRMs,
both in the symmetric case ($\epsilon =0$)  and the asymmetric case ($%
\epsilon \neq 0$).  Near the EPs $g=0.6828$ for $\epsilon =0.2$ and $g=0.3358$ for $\epsilon =0.4$, the QFI in the PTQRM exhibits a sharp peak. This suggests  that the $\mathcal{PT}$-symmetric non-Hermitian system is highly sensitive to perturbations in the estimated parameter near the EP.  This effect is known as $\mathcal{PT}$-enhanced quantum sensing ~\cite{yu2020,ding2023,wang2024}. In the Hermitian QRM, the QFI shows a  shallow maximum. The Hermitian QRM shows only a crossover, without a phase
transition exhibiting a singularity. In contrast, the EPs of the PTQRM mark
the transition from the $\mathcal{PT}$-unbroken phase to the $%
\mathcal{P}\mathcal{T}$-broken phase. Figure~\ref{QCE} (d) shows that  the
QFI in the fourth excited state near the first EP is significantly  higher
than that in the ground state for the same model parameters. This implies
that using the higher excited state can improve the estimation accuracy of unknown parameters.

Since the PTQRM can be viewed as a NHTLS coupled to
a quantized cavity, a natural question arises: Can quantum sensing for the
NHTLS be enhanced through coupling with the cavity. To address this issue,
we calculate the QFI for both the NHTLS and PTQRM, where the former is a component of the latter.  To provide a clear comparison, we derive the analytical QFI for the NHTLS and PTQRM in the AA below,	
\begin{equation}
F_{\epsilon }^{NHTLS}=\left\{
\begin{array}{lll}
\frac{1}{4\left( \Delta ^{2}-\epsilon ^{2}\right) }, & \epsilon \leqslant  \Delta \\
\frac{\Delta ^{2}}{4\epsilon ^{2}\left( \epsilon ^{2}-\Delta ^{2}\right) },
& \epsilon >  \Delta
\end{array}%
\right.   \label{QFI_TLS}
\end{equation}%
and
\begin{equation}
F_{\epsilon }^{PTQRM}=\left\{
\begin{array}{lll}
\frac{1}{4\left( \Delta ^{2}e^{-4g^{2}}-\epsilon ^{2}\right) }, & \epsilon\leqslant \Delta
e^{-2g^{2}} \\
\frac{\Delta ^{2}-g^{2}}{4\epsilon ^{2}\left( \epsilon ^{2}-\Delta
^{2}e^{-4g^{2}}\right) }, & \epsilon >  \Delta  e^{-2g^{2}}
\end{array}%
\right.   \label{QFI_PTQRM}
\end{equation}
where EPs are given by $\epsilon_{EP}=\Delta$ for the NHTLS and $\Delta e^{-2g^{2}}$ for the PTQRM.

In the $\mathcal{PT}$-unbroken regime ($\epsilon\leqslant\epsilon_{EP}$), let bias $\epsilon$ deviate slightly from the EP by a small amount $\delta$; thus, the QFI becomes
\begin{equation*}
F_{\epsilon =\Delta e^{-2g^{2}}-\delta }^{PTQRM}=\left( 8\Delta \delta
\right) ^{-1}e^{2g^{2}}
\end{equation*}%
for the PTQRM in the AA, and
\begin{equation*}
F_{\epsilon =\Delta -\delta }^{NHTLS}=\left( 8\Delta \delta \right) ^{-1},
\end{equation*}%
for the NHTLS. Here, higher-order terms in small $\delta$  are neglected. The QFI is finite as long as  $\delta\ne 0$.  It is evident that the QFI in the PTQRM is higher than that in the NHTLS, with an enhanced factor of $e^{2g^{2}}$. At weak coupling, the enhanced factor can be neglected; however, it becomes more pronounced as the coupling strength increases.	Clearly, these findings also apply to the  $\mathcal{PT}$-broken regime.

{Figure~\ref{QCE}(b) shows the difference in QFIs between the PTQRM and the NHTLS in the ground state as a function of coupling strength $g$ and bias $\epsilon$. Both the PTQRM and NHTLS show a significant enhancement near the EPs. The numerically exact QFI (solid colored line) and the analytical QFI in the AA (dashed line) for the PTQRM are shown in Fig.~\ref{QCE}(d), along with the QFI for the NHTLS ( solid black line) for comparison.  The bias value is normalized with respect to  the corresponding  EPs.} 	At weak coupling ($g=0.1$), the QFI is nearly unaffected by the cavity coupling but increases as the coupling strength increases.   In the deep strong-coupling regime ($g=1.0$), the QFI is substantially  higher than that in the NHTLS.  Strong qubit-cavity coupling
offers an alternative approach  to enhance quantum sensing in the extensively studied NHTLS. Note that the QFI calculated by the AA is quite accurate because only the same eigenstates are used.

\begin{figure}[tbph]
\includegraphics[width=\linewidth]{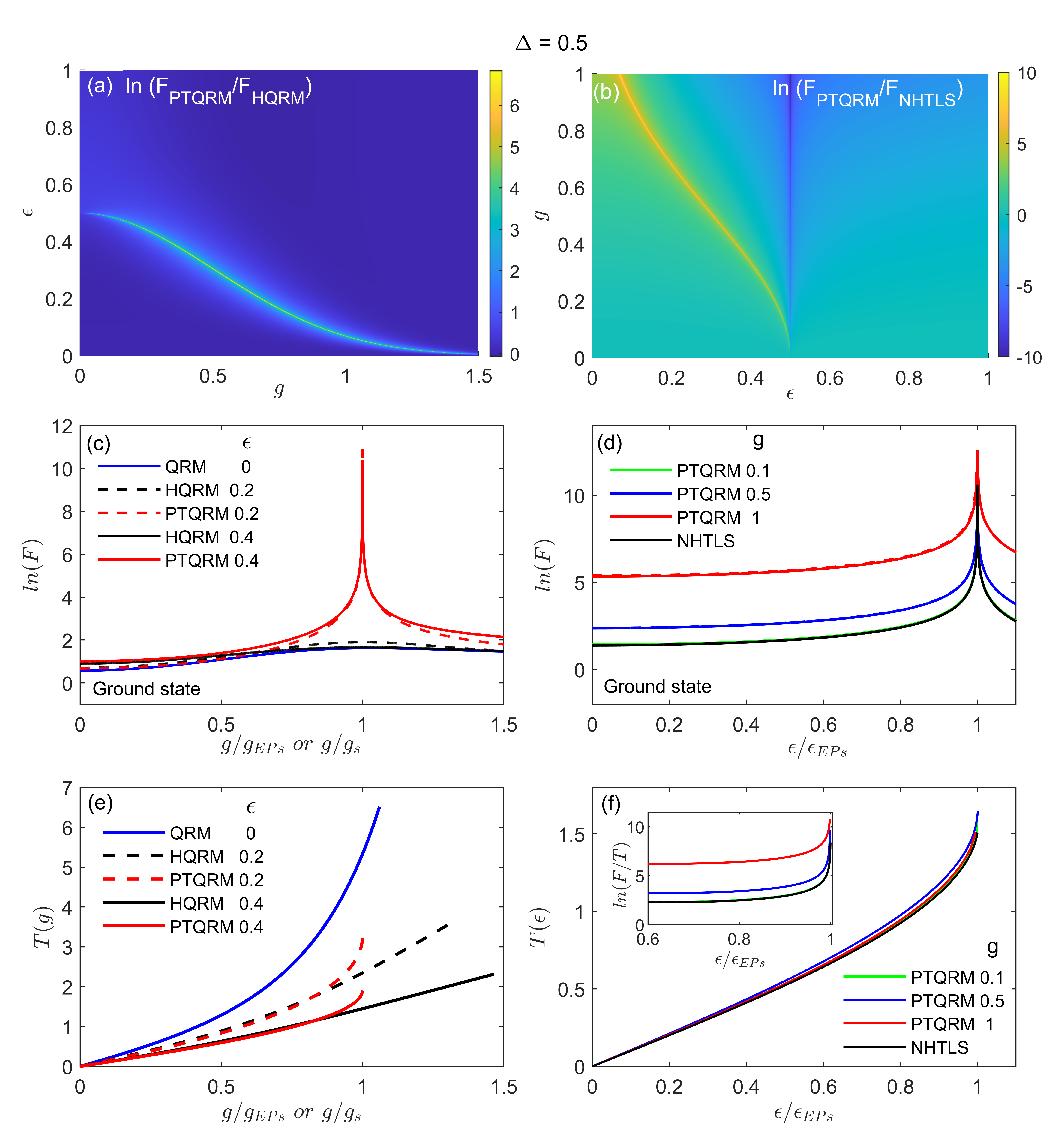}
\caption{
{ Top row: (a) The difference in QFIs  between the PTQRM and the Hermitian QRM, and (b) the difference between the PTQRM and the NHTLS,  as a function of coupling strength $g$ and bias $\epsilon$ in the ground state.} Middle row: (c) The QFI as a function of coupling strength $g$ in the ground state, with $\epsilon = 0.2$ (dashed line) and $\protect\epsilon = 0.4$ (solid line). The black line represents  the Hermitian QRM, the red line represents  the PTQRM, and the blue line represents  the symmetric QRM. {The QFI is scaled by $g/g_{EP}$ or $g/g_{s}$, where $g_{EP}$ is the corresponding exact value at the EP and the $g_{s}$ is the exact value for the  crossover in the Hermitian QRM.} (d) The QFI as a function of bias $\epsilon$ in the ground state for $g$ = 0.1 (green line), 0.5 (blue line), and 1 (red line) in the PTQRM, and for NHTLS (black line). Dashed lines represent the corresponding analytical results for the PTQRM [Eq. ~(\ref{QFI_PTQRM})] and NHTLS [Eq. ~(\ref{QFI_TLS})]. The QFI is scaled by $\protect\epsilon/\epsilon_{F_{max}}$, where $\epsilon_{EP}$ is the  exact value at the EP. { Bottom row: Time $T$ required to prepare the ground state of the systems: (e) for different biases  $\epsilon$ and (f)  for different coupling strengths  $g$. The legends of (e) and (f) are the same as those for  (c) and  (d),  respectively.   { The inset of (f) gives the QFI over the preparation time as a function of normalized {biases} for different coupling strengths .} } 
}
\label{QCE}
\end{figure}

Although a peak appears  in the QFI near the EPs, a higher QFI does not imply directly better metrological  performances. Here, we also evaluate the time required to reach the ground  state toward the EPs of the PTQRM. Critical slowing down  also occurs in the non-Hermitian systems, where the time required to prepare the ground state becomes
extremely slow as it approaches the EPs, similar to open and closed quantum Rabi models~\cite{garbe2020a,montenegro2021,ying2022c}. We consider an adiabatic sweep where the ground state  reaches the equilibrium state,  representing the static approach in critical quantum metrology. We  compare the time $T$ required  to prepare the ground state  among the PTQRM, the Hermitian QRM (HQRM), and the NHTLS. We calculate the preparation time using the bound $T\left(\lambda_c\right) \geq \int_0^{\lambda_c} \frac{1}{\tilde{\Delta}(\lambda)} d \lambda$  to calculate the preparation time,  based on the evolution time $T=\int_0^\lambda \frac{1}{v(\lambda)} d \lambda$, where  $v⁡(\lambda)=d\lambda/dt$ is a speed at which the parameter $\lambda$ evolves  during the adaptative process\cite{garbe2020a}.  The condition $d \lambda / d t \ll \tilde{\Delta}(\lambda)$ holds, where the $\tilde{\Delta}(\lambda)$ is the energy gap between the ground state and first excited state\cite{garbe2020a}.  In adiabatic evolution, the time required to reach the ground state is typically inversely  related to the energy gap $\tilde{\Delta}(\lambda)$. As the gap closes at the EPs,   the dynamical timescale $\tilde{\Delta}(\lambda)^{-1}$ diverges, causing  the time required to prepare the ground state to become extremely slow. Figure 8(e) shows that the preparation time in the PTQRM is generally longer than the H-QRM. As  the bias $\epsilon$ increases, which corresponds to a higher QFI in the PTQRM, the preparation time also increase. Figure 8(f) shows that the preparation time in the PTQRM is slightly longer than  in the NHTLS, but a much higher QFI can be achieved by controlling the coupling strength $g$ in the PTQRM. As the  non-Hermitian system approaches the EPs, the energy gap becomes very small but nonzero, and the QFI increases significantly. This implies that,  despite the longer preparation time, the sensitivity of the system is still enhanced. {The quantity $F/T$ is shown in the inset of Fig. 8(f).	Near the EPs, the QFI increases more rapidly than the preparation time, offering an advantage in efficiently achieving higher QFI and enhanced sensitivity before reaching the EPs.		}	 
 
Finally, the QFI can serve as a signature for the EPs. This indicates that the quantum-criticality-enhanced effect can locate the EPs of the PTQRM, offering an alternative to the $G$ function technique discussed in Sec. II A. 
 Furthermore,the QFI for both the ground state (cf. Fig.~\ref{QCE} ) and the excited states (not shown here) display sharp peaks exactly at the corresponding EPs.

\section{Conclusion}

We use the Bogoliubov operator approach to derive the $G$ function for the PTQRM, whose zeros exactly determine the eigenvalues in both  the
$\mathcal{PT}$-symmetric  and $\mathcal{PT}$-broken regimes.  We also derive the real form of the $G$ function, which accounts for the real eigenvalues in the $\mathcal{PT}$-unbroken phase.	Using this real $G$ function and its first-order derivative, we can analytically detect the EPs, at which both the $G$ function and its first-order derivative with respect to the eigenvalues are simultaneously zero. The previous AA method can be formulated within the current BOA scheme by considering the transitions between the same manifolds. Furthermore, we  improve the AA method by additionally incorporating  the transitions among the nearest-neighboring manifolds. 	Compared to the exact solutions, the present CAA  improve the  eigenvalues and, more importantly, the eigenstates over the AA   in both  the $\mathcal{PT}$-symmetric and $\mathcal{PT}$-broken regimes.

We then study the VRS and atomic dynamics using the CAA in the PTQRM in various coupling regimes and different imaginary biases. The emission spectrum shows two main peaks with unequal height in the ulrastrong-coupling regime. The VRS becomes of multiple Rabi frequencies with increased coupling strength and  higher biases. The width of the peaks also increase with the  coupling strength and the imaginary biases, reflecting the nature of the open quantum systems, such as  the atom-cavity coupling systems dissipating in the environment. This is due to the non-Hermitian description inheriting the nature of the Markov dynamics of the dissipative systems. Interestingly, the CAA results agree well with  the exact solution in all cases, indicating the CAA, where  the transitions within the same manifold and between the nearest-neighboring manifolds are considered,   captures the main features of the atomic dynamics  and emission spectrum in this non-Hermitian light-matter interaction system. In contrast, the AA completely fail to describe the dynamics of the  system, indicating the transition between the same manifold is insufficient to describe the dynamics. {The predicted VRS can be observed experimentally by realizing this model.}

We finally calculate the QFI of the PTQRM, which indicates that quantum criticality-enhanced effects also exist in this non-Hermitian system.	Interestingly, near the EPs, the QFI in the PTQRM is much higher than that in its Hermitian counterpart.	More interestingly, compared to the NHTLS, the QFI is significantly enhanced in the PTQRM in the deep-strong coupling regime. The latter is just a NHTLS coupled with a cavity, indicating that the atom-cavity coupling can enhance the sensitivity, providing a  way to increase the estimation precision.

In summary, we have proposed analytical methods to solve a class of many-body non-Hermitian quantum systems, bridging the gap between Hermitian and non-Hermitian systems, as well as between simple and complex non-Hermitian systems. We expect that the VRS could also be  observed in these non-Hermitian atom-cavity coupling systems. We show that a simple non-Hermitian system, further coupled with a new degree of freedom, provides a  avenue to enhance  quantum sensitivity, serving as a cornerstone for future exploration.	 { Future research should investigate  quantum sensing using the dynamical approach, as it may provide  an alternative pathway for protocol optimization, circumventing the strict dependence on precise state preparation or adiabatic conditions~\cite{chu2021}. } 

\begin{acknowledgments}
This work is supported by the National Key R$\&$D  Program of China (Grant No.  2024YFA1408900) and the National Science Foundation of China (Grant No. 12305032).
\end{acknowledgments}

\begin{figure}[tbph]
\includegraphics[width=\linewidth]{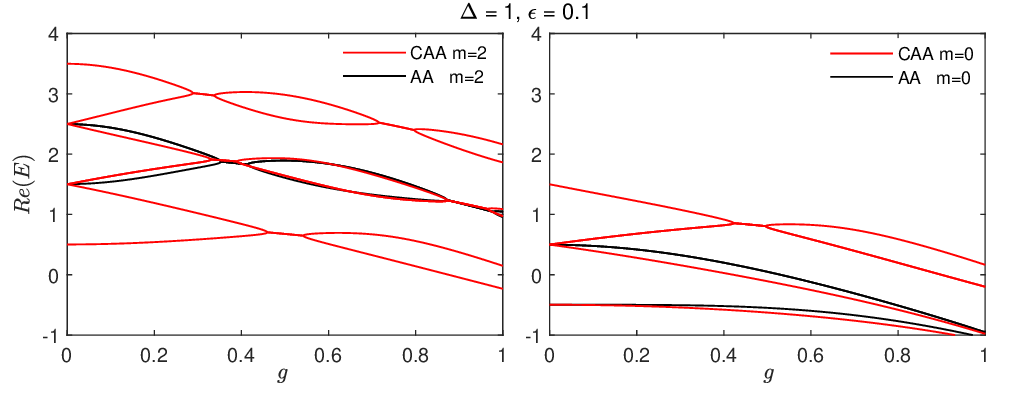}
\caption{ The real part of the eigenvalues given by CAA and
AA as a function of coupling strength $g$ at resonance ($\Delta= 1$), with $\protect\epsilon = 0.1$: (left) results for $m=2$ and  (right) results for $m=0$. The black lines represent AA results, while the red lines denote CAA results.}
\label{spectrum_CAA_6}
\end{figure}
\begin{appendix}

\section*{Appendix:  CAA METHOD}

In the CAA, the  $m$-th pair of eigenvalues$\ $can be obtained by select the
following determinant in a $6\times 6$ block form

\begin{widetext}
\begin{equation}
\begin{vmatrix}
		\Omega_{m-1}^{+}(E) & 0 &0 &	-D_{m-1,m-1} &  -D_{m-1,m} &-D_{m-1,m+1} \\
		0 & \Omega_{m}^{+}(E) & 0 & -D_{m-1,m} &  -D_{m,m} & -D_{m,m+1} \\
		0 & 0 & \Omega_{m+1}^{+}(E) & -D_{m-1,m+1} &  -D_{m,m+1} &-D_{m+1,m+1} \\
		-D_{m-1,m-1} &  -D_{m-1,m} &-D_{m-1,m+1} & \Omega_{m-1}^{-}(E) & 0 & 0\\
		-D_{m-1,m} &  -D_{m,m} & -D_{m,m+1} & 0 & \Omega_{m}^{-}(E) & 0 \\
		-D_{m-1,m+1} &  -D_{m,m+1} &-D_{m+1,m+1} & 0 & 0 & \Omega_{m+1}^{-}(E)
\end{vmatrix} = 0.
\end{equation}
\end{widetext}

Note that for $m=0$, the corresponding determinant takes a $4\times 4$ block
form, because the $m=-1$ manifold does not exist. The determinant yields six
roots for $m>0$ and four roots for $m=0$, but only two are required for any $m$. As we
mentioned in the main text, the CAA eigenvalues should match the AA
eigenvalues.
As a detailed example, Fig. ~\ref{spectrum_CAA_6} shows six (four) groups of the real roots for $m=2(m=0)$ in the left-hand (right-hand) panel.
Two solutions for real eigenvalues in the AA are also collected. One can
immediately see that the third and the fourth roots (the first and the
second roots) of the determinant fit the corresponding real eigenvalues
levels of the AA for $m>0 (m=0)$, while the other solutions are far from
the AA results. This selection rule applies to arbitrary $m$. The
corresponding imaginary parts of these two roots can also be calculated
simultaneously.

\end{appendix}

\bibliographystyle{apsrev4-2}

%\bibliography{NHbias}

%

\end{document}